\documentclass[conference, a4paper]{IEEEtran}
\IEEEoverridecommandlockouts
\usepackage{cite}
\usepackage{amsmath,amssymb,amsfonts}
\usepackage{graphicx}
\usepackage{textcomp}
\usepackage{xcolor}
\usepackage{multirow}
\usepackage{booktabs}
\usepackage{bbold}
\usepackage{url}
\usepackage{siunitx}
\usepackage{subcaption}
\usepackage{cuted}
\usepackage[caption=false]{subfig}
\usepackage[nolist]{acronym}
\usepackage[inline]{enumitem}
\usepackage{optidef}

\usepackage{algorithm}
\usepackage{algorithmicx}
\usepackage[noend]{algpseudocode}

\renewcommand{\baselinestretch}{0.89}
\usepackage[margin=0.72in]{geometry}       
\begin{document}

\makeatletter

\title{Scalable and Efficient Aggregation of Energy-Constrained Flexible Loads 
\thanks{This work was supported as a part of NCCR Automation, a National Centre of Competence in Research, funded by the Swiss National Science Foundation (grant number 51NF40\_225155).}
}
\author{\IEEEauthorblockN{Julie Rousseau\IEEEauthorrefmark{1}\IEEEauthorrefmark{3}\IEEEauthorrefmark{2},
Philipp Heer\IEEEauthorrefmark{3},
Kristina Orehounig\IEEEauthorrefmark{4},
Gabriela Hug\IEEEauthorrefmark{1}}
\IEEEauthorblockA{\IEEEauthorrefmark{1}Power Systems Laboratory, ETH Zürich, Zürich, Switzerland}
\IEEEauthorblockA{\IEEEauthorrefmark{3} Urban Energy Systems Laboratory, Empa, D\"{u}bendorf, Switzerland}
\IEEEauthorblockA{\IEEEauthorrefmark{4} Research unit for Building Physics and Building Ecology, Vienna University of Technology, Vienna, Austria}
\IEEEauthorblockA{\IEEEauthorrefmark{2}jrousseau@ethz.ch}
}

\maketitle

\begin{abstract}
    Loads represent a promising flexibility source to support the integration of renewable energy sources, as they may shift their energy consumption over time. By computing the aggregated flexibility of power and energy-constrained loads, aggregators can communicate the group's flexibility without sharing individual private information. However, this computation is, in practice, challenging. Some studies suggest different inner approximations of aggregated flexibility polytopes, but all suffer from large computational costs for realistic load numbers and horizon lengths. In this paper, we develop a novel approximation of the aggregated flexibility of loads based on the concept of worst-case energy dispatch, i.e., if aggregated energy consumptions are assumed to be dispatched in the worst manner possible. This leads to conservative piecewise linear bounds that restrict the aggregated energy consumption only based on the previous aggregated energy consumed. A comparative case study reveals that our method can compute an approximation of the aggregation of thousands of loads efficiently, while displaying an accuracy comparable to other approximation techniques.  
\end{abstract}


\begin{acronym}[ML] 
    \acro{DTU}{Technical University of Denmark}
    \acro{EV}{Electric Vehicle}
    \acro{HVAC}{Heating, Ventilation and Air Conditioning}
\end{acronym}

\section{Introduction}

In an attempt to reduce global carbon emissions, electric power grids undergo profound transformations: a large number of flexible carbon-intensive power plants are planned to be decommissioned and replaced by intermittent renewable energy sources \cite{ReportIRENA2019}. While conventional power plants used to adapt their production to the electric power demand, renewable power units can only partly do so. 
Hence, policymakers encourage power consumers to become more flexible, i.e., partly adapt their consumption to the production of renewable units \cite{EUFlexibility}.

Many devices, referred to as loads in this paper, can consume power flexibly. In particular, \acp{EV} and \ac{HVAC} systems enable a large flexibility potential.
Yet, these flexible loads are distributed across power grids, making their integration into network operations challenging. 
Aggregators can support such integration while also addressing privacy and computational concerns \cite{Eid2015}. This intermediary gathers consumers' private information and communicates aggregated power constraints to system operators, reducing the communication burden on system operators and masking individual private information. 

Computing aggregated flexibility constraints is a challenging task for aggregators. Indeed, most flexible loads are power and energy-constrained. Hence, their feasible power consumption region can be described as a polytope, i.e., a convex set delimited by linear constraints. The Minkowski sum of individual polytopes defines the feasible aggregated power, i.e., the range of aggregated power that can be decomposed into individual feasible power. However, the Minkowski sum's computation time grows exponentially with the aggregated load number \cite{Trangbaek2011}. 

Hence, some studies propose more efficient Minkowski sum computation techniques, exploiting the properties of flexible load polytopes. The aggregated polytope's linear constraints can be computed more efficiently using constraint reduction methods \cite{Trangbaek2012, Wen2022}. Alternatively, a tree-search based on extreme load behaviors allows a more efficient vertices' computation \cite{Trangbaek2011, Ozturk2024}. While such methods accelerate the aggregated polytope's computation, they remain inapplicable for large numbers of loads due to computational costs. Some simplifications, e.g., the computation of only a few aggregated vertices, may reduce computational costs at the expense of accuracy \cite{Ozturk2024}. Another research direction focuses on geometric inner approximations of the aggregated polytope. Specific geometric shapes, e.g., homothets \cite{Zhao2017} or zonotopes \cite{Mueller2015}, can provide an inner approximation of individual load polytopes so as to compute their Minkowski sum efficiently, exploiting the selected shape's properties. The aggregated polytope can also be approximated by a specific shape, e.g., the largest inscribed homothet \cite{Zhao2016} or ellipsoid \cite{Zhen2018}. However, all inner approximation methods require finding the largest inscribed geometric shape in a polytope with the help of a convex optimization, leading to significant computational costs. A comparative study reveals that such approximations suffer from too large computational costs for realistic load numbers and horizon lengths \cite{OzturkToolbox2022}. A recent contribution investigates the concept of worst-case power trajectories to describe the aggregated energy flexibility of loads. Only feasible aggregated energy values that are feasible for all possible power trajectories are considered \cite{PANDA2024}. However, the method generates a constraint number that grows exponentially with the horizon length, leading to an unacceptable communication and computational burden in practice. 

To the best of our knowledge and as highlighted in \cite{OzturkToolbox2022}, no existing method can approximate the loads' aggregated flexibility efficiently and accurately for realistically large load numbers and horizon lengths, i.e., thousands of loads and up to one day ahead. In this paper, we develop a novel approximation of the aggregated flexibility of energy and power-constrained loads, based on load polytope properties and the concept of worst-case energy dispatch. With a comparative case study, we demonstrate that, unlike other approximations, our method displays acceptable computational performances for realistic load aggregations.   

The remainder of this paper is organized as follows. Section~\ref{sec:load} describes how power and energy-constrained loads are represented and explores the properties of their aggregation. Exploiting such characteristics, Section~\ref{sec:aggregation} explains the novel approximation of loads' aggregated flexibility developed in this paper. Section~\ref{sec:caseStudy} introduces the case study and Section~\ref{sec:Results} presents the results and discussion. Finally, Section~\ref{sec:Conclusion} concludes the work.

In the remainder of this paper, bold letters designate vectors or matrices. The sets $\mathcal{T}$ and  $\mathcal{I}$ describe the discrete set of loads and timesteps, respectively. $T$, $N$, and $\Delta t$ consistently designate the number of timesteps, the number of loads, and the time resolution, respectively.

\section{Energy-Constrained Loads}
\label{sec:load}
This section introduces the representation of flexible loads adopted in this paper and studies properties of their aggregation. 

\subsection{Individual Loads' Polytope}

\subsubsection{Power Polytope}

The power consumption of a flexible load $i$ must satisfy its power constraints: 
\begin{equation}
    \underline{p}_i \leq p_{i,k} \leq \overline{p}_i, \quad \forall k \in \mathcal{T},
    \label{eq:powerConstraint}
\end{equation}
where $\underline{p}_i$ and $\overline{p}_i$ denote the minimum and maximum load's power ratings. Additionally, its energy content must satisfy: 
\begin{equation}
    \underline{e}_{i,k} \leq \Delta t \sum_{l = 0}^k p_{i,l} \leq \overline{e}_{i,k}, \quad \forall k \in \mathcal{T},
    \label{eq:energyConstraint}
\end{equation}
where $\underline{\pmb{e}}_i$ and $\overline{\pmb{e}}_i$ denote the load's time-varying energy limits. A methodology to compute energy bounds of \acp{EV} and \ac{HVAC} systems can be found in \cite{GASSER2021}. The power and energy constraints limiting the power consumption of flexible load~$i$ form the load's power polytope: 
\begin{equation}
    \mathcal{P}_i^p  = \left\{ \pmb{p} \in \mathbb{R}^T | \mathbf{A}_i^p \pmb{p} \leq \pmb{b}_i^p \right\},
\end{equation}
where $\mathbf{A}_i^p$ and $\pmb{b}_i^p$ are derived from (\ref{eq:powerConstraint}) and (\ref{eq:energyConstraint}).

\subsubsection{Energy Polytope}
Other load aggregation studies formulate load constraints in terms of power. Instead, in this paper, we express load constraints in terms of energy. The power constraints of load $i$ become: 
\begin{equation}
\begin{aligned}
    \underline{p}_i & \leq  \frac{e_{i,k} - e_{i,k-1}}{\Delta t} \leq \overline{p}_i, \quad \forall k \in \mathcal{T} \setminus \{ 0\}, \\
    \underline{p}_i & \leq \frac{e_{i,0} }{\Delta t} \leq \overline{p}_i, 
\end{aligned}
\label{eq:powerConstraintsEnergy}
\end{equation}
and its energy constraints are formulated as: 
\begin{equation}
     \underline{e}_{i,k} \leq e_{i,k} \leq \overline{e}_{i,k}, \quad \forall k \in \mathcal{T}.
     \label{eq:energyConstraintsEnergy}
\end{equation}
While a load's power consumption depends on its entire power consumption history, its energy consumption is constrained by its last energy content. That is why we study the load's energy polytope, defined as: 
\begin{equation}
    \mathcal{P}_i^e = \left\{ \pmb{e} \in \mathbb{R}^T | \mathbf{A}_i^e \pmb{e} \leq \pmb{b}_i^e \right\},
    \label{eq:energyPolytope}
\end{equation}
where $\mathbf{A}_i^e$ and $\pmb{b}_i^e$ are derived from (\ref{eq:powerConstraintsEnergy}) and (\ref{eq:energyConstraintsEnergy}). 

\subsubsection{Assumptions} 
For simplicity, we assume constant power ratings over time in this paper, similarly to \cite{PANDA2024} and \cite{Ozturk2024}. While this assumption holds for stationary batteries and \ac{HVAC} systems, it cannot describe \ac{EV} departures. Furthermore, we restrict our analysis to positive power ratings. Nevertheless, we believe that the developed methodology can be extended to time-varying negative power ratings in the future. Additionally, if flexible loads suffer from losses, e.g., \ac{HVAC} systems, conservative energy bounds as introduced in \cite{operativeEnvelope} allow to account for losses while conforming to the formalism of (\ref{eq:powerConstraint}) and (\ref{eq:energyConstraint}).

\subsection{Properties of the Aggregated Loads' Polytope}
\label{sec:properties}

Individual load energy polytopes can be represented in a standardized manner using (\ref{eq:energyPolytope}). Here, we examine the properties of their Minkowski sum in order to better comprehend the complexities associated with aggregation. 

\subsubsection{Load aggregation over a two-timestep horizon}

\begin{figure}
    \centering
    \hspace{-1cm}
    \includegraphics[width=\columnwidth]{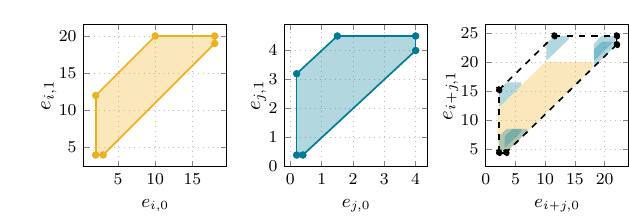}
    \caption{Aggregation of two loads over two timesteps.}
    \label{fig:minkowskiSumExample}
\end{figure}

A flexible load's energy consumption is constrained by (\ref{eq:powerConstraintsEnergy}) and (\ref{eq:energyConstraintsEnergy}). Over a two-timestep horizon, i.e., when $T=1$, its energy polytope is a trapezoid, as illustrated in Fig.~\ref{fig:minkowskiSumExample} for an example of two loads. Hence, it is defined by at most 6~vertices. Furthermore, the two diagonal lines representing the power constraints are parallel, and their slopes, equal to $1/\Delta t$, are identical across all loads. Fig.~\ref{fig:minkowskiSumExample} also shows the aggregation of two loads, $i$ and $j$, i.e., their Minkowski sum. Since all diagonal lines have an identical slope, the aggregated polytope's vertices can be efficiently computed as the algebraic sum of the individual vertices. Therefore, in the case of a two-timestep horizon, the exact aggregated polytope can be computed efficiently. 
                                   
\subsubsection{Time-coupling in the aggregated polytope}

A load's energy polytope is sequential in time, i.e., its energy consumption depends only on its last energy content. Here, we investigate if such property also holds for an aggregation of loads, i.e., if the range of feasible aggregated energy depends only on the last aggregated energy consumption. 

We consider a three-timestep horizon, i.e., $T=2$. As previously explained, the aggregated energy in the first two timesteps is delimited by a trapezoid. Similarly, in the second and third timesteps, the feasible individual loads' energy are contained in trapezoids that can be efficiently aggregated. This second aggregated trapezoid describes the Minkowski sum of load polytopes over the second and third timesteps only. We select an aggregated energy consumption trajectory such that $\left(e_0, e_1\right)$ belongs to the first trapezoid and $\left(e_1, e_2\right)$ to the second. As $\left(e_0, e_1\right)$ belongs to the first trapezoid, the two first aggregated energy values can be disaggregated into individual feasible energy consumptions. Similarly, there exists a feasible disaggregation of $\left(e_1, e_2\right)$. However, the two decompositions of the aggregated energy $e_1$ may be different. The initial decomposition of $e_1$ may even restrict the feasible aggregated energy $e_2$.

We provide an example that confirms this intuition. We consider two batteries, $a$ and $b$, with their maximum power and energy characteristics specified in Table~\ref{tab:example}. Battery $a$ is constrained by its power rating during the three timesteps, while battery $b$ is limited by its energy capacity. Their maximum aggregated power and time-varying aggregated energy are specified in Table~\ref{tab:example}. For instance, the successive aggregated energy consumption values $\pmb{e}_\mathrm{agg} = \left(2, 2, 4 \right)$~kWh respect the aggregated maximum power and energy constraints. To reach the selected $e_0$, battery $a$ and $b$ must both absorb 1~kW. While battery $b$ is already fully charged, it must absorb 1~kW more in the third timestep. This counter-example proves that an aggregated polytope is not sequential in time, i.e., the range of feasible aggregated energy depends on all past energy values and not only the last one. Besides, for a maximum power request, all batteries must consume at their maximum power. Hence, high-power/low-energy batteries are fully charged quickly, reducing the maximum power for the rest of the horizon. Based on this physical understanding, we develop our aggregation method. 

\begin{table}[t]
    \centering
    \footnotesize
    \renewcommand{\arraystretch}{1.3}
    \setlength{\tabcolsep}{0pt}
    \caption{\small Parameters used to illustrate the time coupling in the aggregated polytope.}
    \label{tab:example}

    \begin{minipage}{0.3\columnwidth}
        \centering
        \begin{tabular}{c c c}
            \toprule
             & \hspace{0.1cm} $a$ \hspace{0.1cm} & \hspace{0.1cm} $b$ \hspace{0.1cm} \\ \midrule
            $\overline{p}$ [kW] & 1 & 3 \\
            $\overline{e}$ [kWh] & 3 & 1 \\
            \bottomrule
        \end{tabular}
    \end{minipage}%
    \begin{minipage}{0.4\columnwidth}
        \centering
        \begin{tabular}{c c c c}
            \toprule
            Timestep & \hspace{0.1cm} $\overline{e}_a$ \hspace{0.1cm} & \hspace{0.1cm} $\overline{e}_b$ \hspace{0.1cm} & Agg. $\overline{e}$  \\ 
             & & [kWh] &   \\ \midrule
            $0$ & 1~ & 1 & 2 \\ 
            $1$ & 2 & 1 & 3 \\
            $2$ & 3 & 1 & 4 \\
            \bottomrule
        \end{tabular}
    \end{minipage}%
    \begin{minipage}{0.24\columnwidth}
    \begin{center}
        \hspace{0.2cm}Agg. $\overline{p}$ = 4~kW
    \end{center}
    \end{minipage}
\end{table}

\section{Aggregation of Energy-Constrained Loads: A Worst-Case Energy Dispatch Approach}
\label{sec:aggregation}

As illustrated, even though individual loads' flexibility is sequential in time, their aggregated flexibility is not. Here, we develop a time-sequential approximation of flexible loads' aggregated flexibility, using the concept of worst-case energy dispatch. In other words, we define an approximation, in which the range of feasible aggregated energy only depends on the last aggregated energy consumption value. We first briefly explain our approach before presenting the complete methodology.

\subsection{Description of our Approach}

The developed methodology is based on the observations presented in Section~\ref{sec:properties}. Inspired by the algebraic summation of trapezoids, our methodology relies on the algebraic summation of individual loads' energy upper and lower bounds, defined for all $k$ as: 
\begin{equation}
    \begin{aligned}
        e_{k+1}^\mathrm{up} &= \sum_{i \in \mathcal{I}} e_{i,k+1}^\mathrm{up} = \sum_{i \in \mathcal{I}} \min \Big( \overline{e}_{i,k+1}; e_{i,k} + \Delta t \overline{p}_i \Big),\\
        e_{k+1}^\mathrm{down} &= \sum_{i \in \mathcal{I}} e_{i,k+1}^\mathrm{down} = \sum_{i \in \mathcal{I}} \max \left( \underline{e}_{i,k+1}; e_{i,k} + \Delta t \underline{p}_i \right).
    \end{aligned}
    \label{eq:bound_k_kplus1}
\end{equation}
Nevertheless, these bounds depend on previous individual loads' energy consumption, $e_{i,k}$, and not on the previous aggregated energy $e_k$. Therefore, our aim is to define some conservative bounds $e_{k+1}^\mathrm{up,c} \left(e_k\right)$ and $e_{k+1}^\mathrm{down,c} \left(e_k\right)$, defined as the minimum value of $e_{k+1}^\mathrm{up}$ and the maximum value of $e_{k+1}^\mathrm{down}$ across all possible energy dispatches summing to $e_k$, respectively. In other words, for any dispatch $\pmb{e}_k^d$, summing up to $e_k$, the bounds are defined as: 
\begin{equation}
\begin{aligned}
     \overbrace{\sum_{i \in \mathcal{I}} \min \Big( \overline{e}_{i,k+1}; e_{i,k}^\mathrm{c} + \Delta t \overline{p}_i \Big) }^{ e_{k+1}^\mathrm{up,c} \left( e_k \right) } & \leq \sum_{i \in \mathcal{I}} e_{i,k+1}^\mathrm{up} \left( \pmb{e}_k^d \right), \\
     \underbrace{\sum_{i \in \mathcal{I}} \max \left( \underline{e}_{i,k+1}; e_{i,k}^\mathrm{c} + \Delta t \underline{p}_i \right)}_{e_{k+1}^\mathrm{down,c} \left( e_k \right)}  & \geq \sum_{i \in \mathcal{I}} e_{i,k+1}^\mathrm{down} \left( \pmb{e}_k^d \right).
\end{aligned}
\label{eq:wc_bounds_1}
\end{equation}
Then, if some aggregated energy values $\left( e_k \right)_{k \in \mathcal{T}}$ recursively fulfill the conservative bounds, i.e.: 
\begin{equation}
    e_{k+1}^\mathrm{down,c} \left( {e}_k \right) \leq e_{k+1} \leq e_{k+1}^\mathrm{up,c} \left( {e}_k \right), \quad \forall k \in \mathcal{T},
\end{equation}
they are also upper and lower bounded by $e_{k+1}^\mathrm{up} \left(\pmb{e}_k^d\right)$ and $e_{k+1}^\mathrm{down} \left(\pmb{e}_k^d\right)$, respectively, for any dispatch $\pmb{e}_k^d$ summing to $e_k$. Consequently, there exists $\lambda \in [0, 1]$, such that:
\begin{equation}
\begin{aligned}
    e_{k+1} & = \lambda e_{k+1}^\mathrm{down} \left( \pmb{e}_k^d \right) + \left( 1 - \lambda \right) e_{k+1}^\mathrm{up} \left( \pmb{e}_k^d \right), \\
    & = \sum_{i \in \mathcal{I}} \left( \lambda e_{i,k+1}^\mathrm{down} \left( e_{i,k}^d \right) + \left( 1 - \lambda \right) e_{i,k+1}^\mathrm{up} \left( e_{i,k}^d \right) \right),
\end{aligned}
\end{equation}
which describes a feasible decomposition.
In other words, if some aggregated energy values lie within the conservative bounds, they can be decomposed into feasible individual energy values, i.e., the conservative bounds define an inner approximation of the aggregated polytope. Furthermore, the set delimited by these bounds is sequential in time.  

\subsection{Worst-Case Energy Dispatch}

Our methodology relies on the definition of the worst-case energy dispatch, $e_{i,k}^\mathrm{c}$, for an aggregated energy value $e_k$.  According to the example introduced in Section~\ref{sec:properties}, it seems that the worst-case energy dispatch consists of extreme power consumption, alternating between maximum and minimum power consumption, which we confirm here.

\subsubsection{Definition}
For an aggregated energy value $e_k$, the worst-case energy dispatch minimizes $e_{k+1}^\mathrm{up}$. Such minimum is obtained when the energy consumption of as many loads as possible is constrained by their maximum energy content at time $k+1$. The worst-case energy dispatch also maximizes $e_{k+1}^\mathrm{down}$, i.e., the energy consumption of as many as loads as possible is constrained by their minimum energy content at time $k+1$. In practice, an aggregator would dispatch energy, such as to minimize the number of loads reaching their minimum and maximum energy contents. Yet, as illustrated in Section~\ref{sec:properties}, for maximum or minimum power requests, the aggregator cannot decide on the loads' dispatch, as they all must consume at their maximum or minimum power, respectively. Therefore, the worst-case energy dispatch consists of successive maximum and minimum power consumption periods, bringing loads close to their maximum and minimum energy contents.

\subsubsection{Computation}
\label{subsubsec:discreteEnergy}
Over the $k$ first timesteps, a load $i$ can consume its maximum power $\overline{p}_i$ over a duration $d_{i,k}$ before reaching its maximum energy content $\overline{e}_{i,k}$:
\begin{equation}
    \overline{e}_{i,k} = d_{i,k} \overline{p}_i \implies d_{i,k} = \frac{\overline{e}_{i,k}}{\overline{p}_i}.
\end{equation}
All loads can consume their maximum power over the shortest computed duration. Then, the load associated with the shortest duration has reached its maximum energy content, while the others keep on consuming their maximum power. Hence, we order the computed durations in ascending order, a process defined by the permutation $\phi_k$, and then compute the discrete energy levels: 
\begin{equation}
    e_k^{\mathrm{tot},j} = \sum_{\phi_k \left( i \right) \leq j} \overline{e}_{i,k} + d_{j,k} \sum_{\phi_k \left( i \right) > j}  \overline{p}_{i}.
\end{equation}
Here, $e_k^{\mathrm{tot},j}$ designates the aggregated energy after which $j$ loads have reached their maximum energy content, in the worst-case energy dispatch. If, for an aggregated energy $e_k$, it holds that: 
\begin{equation}
    e_k^{\mathrm{tot},j} \leq e_k < e_k^{\mathrm{tot},j+1},
\end{equation}
the associated worst-case energy dispatch of load $i$ is:
\begin{equation}
    e_{i,k}^\mathrm{c} \left( e_k \right) = 
    \begin{cases}
        \overline{e}_{i,k}, & \text{if } \phi_k \left(i\right) \leq j, \\
        \left( d_{j,k} + \frac{e_k - e_k^{\mathrm{tot},j}}{\sum_{\phi_k \left(l \right) > j} \overline{p}_{l}} \right) \overline{p}_i, & \text{else}. 
    \end{cases}
    \label{eq:worstCaseDispatch}
\end{equation}
Hence, the worst-case energy dispatch of $i$ is a piecewise linear function of $e_k$, composed of at most $N$ segments.

\subsection{Worst-Case Energy Bound Computation}

Given the worst-case energy dispatch defined in (\ref{eq:worstCaseDispatch}), we can now determine the conservative energy bounds introduced in~(\ref{eq:wc_bounds_1}). We obtain for an aggregated energy consumption $e_k$ comprised between $e_k^{\mathrm{tot},j}$ and $e_k^{\mathrm{tot},j+1}$: 
\begin{equation}
\begin{aligned}
    e_{k+1}^\mathrm{up,c} \left(e_k \right) = & \sum_{i \in \mathcal{I}} \min \left( \overline{e}_{i,k+1}; e_{i,k}^\mathrm{c} \left( e_k \right) + \Delta t \overline{p}_i \right), \\
    = & \sum_{\phi_k \left(i \right) \leq j} \min \left( \overline{e}_{i,k+1}; \overline{e}_{i,k} + \Delta t \overline{p}_i \right) \\
    + & \sum_{\phi_k \left(i \right) > j} \min \left(\overline{e}_{i,k+1}; {e}_{i,k}^\mathrm{c} \left( e_k \right) + \Delta t \overline{p}_i \right)
\end{aligned}
\end{equation}
Given $e_k$, $j$ loads are fully charged in the worst-case energy dispatch, and the dispatch of the remaining loads depends on $e_k$. However, the remaining loads may become fully charged at timestep $k+1$, and their contribution to the conservative upper bound may transition from linearly dependent on $e_k$ to constant. Therefore, we must compute up to $N-j$ additional discrete energy values at timestep~$k$ to fully characterize the piecewise linear relationship for $e_{k+1}^\mathrm{up,c} \left(e_k \right)$. All in all, the conservative upper energy bound is a piecewise linear function of $e_k$, characterized by at most $N \left( N+3 \right) / 2$ segments. Using a similar procedure, we can compute the piecewise linear relationship between the lower conservative energy bound $e_{k+1}^\mathrm{down,c}$ and $e_k$.

\subsection{Linear Approximation of the Worst-Case Energy Bounds}
\label{subsec:linearApp}

The proposed piecewise linear approximation presents two practical limitations. First, the approximation may be composed of a large number of segments, creating many constraints on the aggregated energy consumption. In most cases, it even generates more constraints than in the unaggregated case, i.e., with individual loads' constraints. Additionally, the approximation may be non-convex, complicating its integration into centralized optimizations. 

Therefore, we suggest linear approximations of the piecewise linear conservative bounds. Specifically, we determine a conservative linear approximation for which the upper (resp. lower) bound lies below (resp. above) all upper (resp. lower) constraints and which maximizes the area of the aggregation's approximation. In practice, at timestep~$k$, given a range of slopes, we analytically compute optimal intercept values and select the lower and upper linear approximations yielding the largest area. If energy values satisfy the linear constraints, they satisfy the piecewise linear constraint and are feasible aggregated energy values.

\section{Case Study and Metrics}
\label{sec:caseStudy}

To assess the developed method, we should not only evaluate its performances but also compare them to existing flexibility aggregation approximations. To this end, we leverage the open-source work developed in \cite{OzturkToolbox2022}, a Python toolbox comparing various inner approximations of the aggregated flexibility of loads. In this section, we introduce the case study, the considered benchmark aggregation methods and the performance metrics. 

\subsection{Case Study}
We consider a set of households with energy-constrained flexible loads. More precisely, each residential household displays an inflexible power consumption and is equipped with a static \ac{EV}, i.e., we assume that the \ac{EV} remains plugged in over the entire horizon. The \acp{EV} are assumed to be plugged to slow home-charging devices, e.g., Tesla Powerwalls \cite{TeslaSpec}. Their maximum power and energy characteristics are uniformly sampled from the ranges specified in Table~\ref{tab:parametersToolbox} to ensure diversity among loads. All \acp{EV}' minimum power are set to 0, i.e., only unidirectional charging is permitted. Additionally, each battery must satisfy a minimum energy requirement at the end of the simulation horizon, sampled from the range indicated in Table~\ref{tab:parametersToolbox}. To consider variability in loads, we consistently evaluate methods on 10~random load groups. 

\begin{table}[t]
    \centering
    \footnotesize
    \renewcommand{\arraystretch}{1.3}
    \setlength{\tabcolsep}{1.5pt}
    \caption{\small Range of battery characteristics in case study \cite{OzturkToolbox2022}.}
    \label{tab:parametersToolbox}

    \begin{minipage}{0.45\columnwidth}
        \centering
        \begin{tabular}{c c c}
            \toprule
            \textbf{Parameter} & Range & Unit \\ \midrule
            $\overline{p}$ & [4.0,  6.0] & kW \\
            $\underline{p}$ & 0 & kW \\
            \bottomrule
        \end{tabular}
    \end{minipage}%
    \begin{minipage}{0.45\columnwidth}
        \centering
        \begin{tabular}{c c c}
            \toprule
            \textbf{Parameter} & Range & Unit \\ \midrule
            $\overline{e}$ & [10.5, 13.5] & kWh \\ 
            $\underline{e}$ & [0.0, 10.5]  & kWh \\
            \bottomrule
        \end{tabular}
    \end{minipage}
\end{table}

\subsection{Benchmark Aggregated Flexibility Methods}
\label{subsec:benchmark}
An extensive literature review and comparative study of flexible loads' aggregation methods can be found in \cite{OzturkToolbox2022}. We test our method against three inner approximation techniques in this work. Specifically, we consider the largest homothet \cite{Zhao2017} and ellipsoid \cite{Zhen2018} inscribed in the aggregated polytope, as well as the aggregation of the largest zonotopes inscribed in individual load's polytopes \cite{Mueller2015}. While other aggregation methods are implemented in \cite{OzturkToolbox2022}, the three selected benchmark techniques show the best trade-off of computational complexity and accuracy in preliminary tests.

\subsection{Metrics}

To compare different approximation methods, the authors in \cite{OzturkToolbox2022} suggest using the aggregated polytope's approximations in realistic use-cases. In particular, they assess the performances of different approximations on the minimum total energy cost and peak power when the flexibility is leveraged in a dispatch optimization. Our first metric evaluates the approximations' accuracies, expressed as the increase in total cost and peak power compared to the unaggregated case, i.e., with all loads' constraints in the optimization. As inflexible demand profiles and energy prices are season-dependent, each optimization is performed for 12~days, i.e., one day per month, always starting at midnight and assuming a time resolution of 15~minutes.

While inner approximation computations are usually computationally expensive, they reduce the number of optimization constraints. Therefore, as a second metric, we monitor the total computation time, i.e., the time needed to compute the approximation and solve the optimization. 

Finally, our method relies on the concept of worst-case energy dispatch. Even though we provide intuitive and analytical reasoning, we do not formally prove that the introduced dispatch is the worst. Therefore, for all aggregated energy requests, we assess if the signal can be decomposed into individual feasible energy consumption values using a disaggregation optimization. The root-mean-squared error between the aggregated command and the optimal aggregated consumption is our third metric.

\section{Results and Discussion}
\label{sec:Results}

This section first illustrates our methodology with an example before comparing it to the existing approximation methods introduced in Section~\ref{subsec:benchmark}.

\subsection{Illustrative Example}

Our methodology is a time-sequential inner approximation of an aggregated polytope. As such, the aggregated energy consumption $e_k$ only depends on the previous aggregated energy $e_{k-1}$. Therefore, the range of feasible aggregated energy consumption $e_k$ can be studied in the $\left( e_k, e_{k-1} \right)$-plane. At timestep $k$, piecewise linear conservative upper and lower bounds constrain $e_k$ as a function of $e_{k-1}$. 

\begin{figure}
    \centering
    \hspace{-1cm}
    \includegraphics[width=\columnwidth]{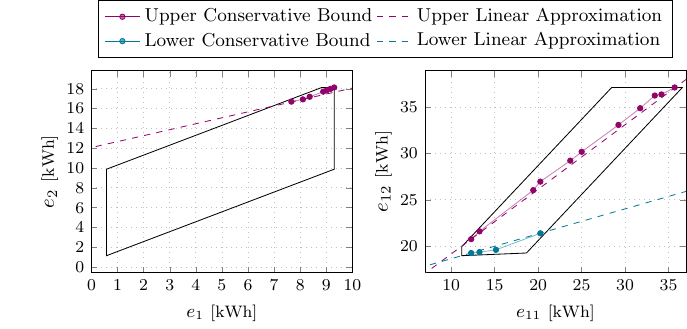}
    \caption{Example of the aggregation of 7 loads over a 3-hour horizon ($T=12$).}
    \vspace{-0.2cm}
\label{fig:illustrationPolytope}
\end{figure}

Fig.~\ref{fig:illustrationPolytope} illustrates our methodology, using the example of 7~aggregated loads over a 3-hour horizon, i.e., a horizon length $T = 12$ for a 15-minute resolution. The conservative upper and lower bounds are depicted with the solid purple and blue lines, respectively. Additionally, the circle markers indicate the discrete energy levels defined in Section~\ref{subsubsec:discreteEnergy}. At the beginning of the horizon, most aggregated energy values $e_1$ cannot lead to fully charged loads, even in the worst-case dispatch. Only large energy values $e_1$ may lead some loads to be fully charged in the worst-case energy dispatch. Therefore, the piecewise linear upper bound only constrains the range of feasible $e_2$ for large values of $e_1$. Furthermore, no feasible aggregated energy value $e_1$ may yield small enough individual energy contents that a conservative lower bound must restrict the range of feasible $e_2$. Towards the end of the horizon, in the $\left( e_{11}, e_{12} \right)$-plane, almost all values of $e_{11}$ may lead to fully charged loads in the worst-case energy dispatch. Therefore, a conservative upper bound constrains the range of feasible $e_{12}$ for all values of $e_{11}$. We also observe a lower conservative bound, restricting $e_{12}$ for small aggregated energy values $e_{11}$. Indeed, for small $e_{11}$ values, the worst-case dispatch may have charged some loads too little, such that they cannot operate at their minimum power anymore, but must instead charge their minimum required energy. 

Fig.~\ref{fig:illustrationPolytope} also illustrates the non-convexity of the piecewise-linear conservative bounds. For instance, the upper conservative bound in the right figure is clearly non-convex. The dashed lines in Fig.~\ref{fig:illustrationPolytope} describe the linear approximations introduced in Section~\ref{subsec:linearApp} to ensure constraint convexity. In this example, linear functions approximate well the upper and lower conservative bounds. However, when aggregating a large number of loads, we find that the lower conservative bound can be better approximated by two linear constraints or a parabolic function, which we leave for future works.

\subsection{Comparative Study}
\begin{figure}
    \centering
    \hspace{-3cm}\includegraphics[width=1.3\columnwidth]{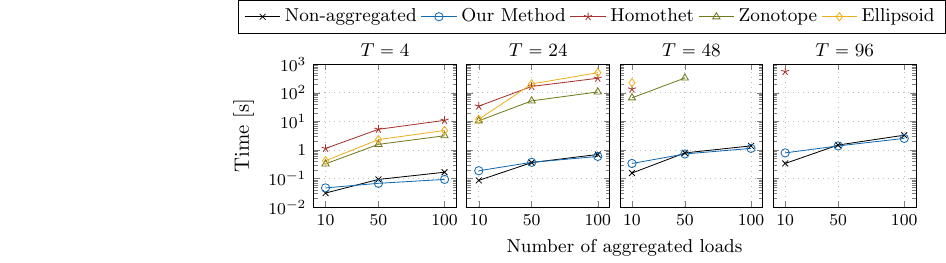}

    \vspace{0.1cm}
    \hspace{-3.2cm}\includegraphics[width=1.25\columnwidth]{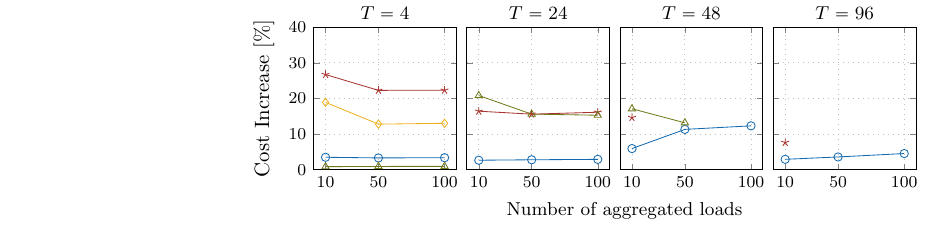}
    \caption{Evolution of the total computation time (top) and the total energy cost increase (bottom) for different horizon lengths and aggregated load numbers.}
    \label{fig:all_methods_cost}
\end{figure}

\begin{figure}
    \centering
    \hspace{-3cm}\includegraphics[width=1.3\columnwidth]{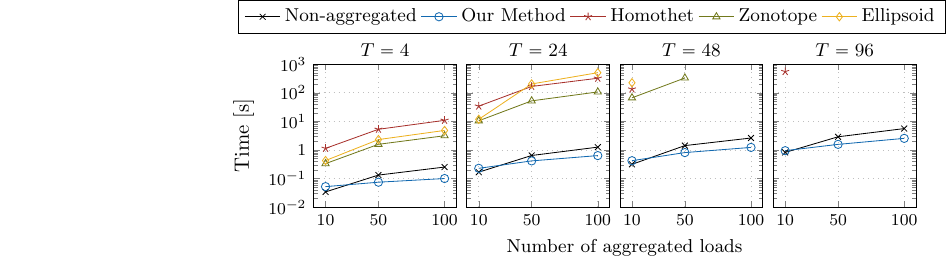}

    \vspace{0.1cm}
    \hspace{-3.2cm}\includegraphics[width=1.25\columnwidth]{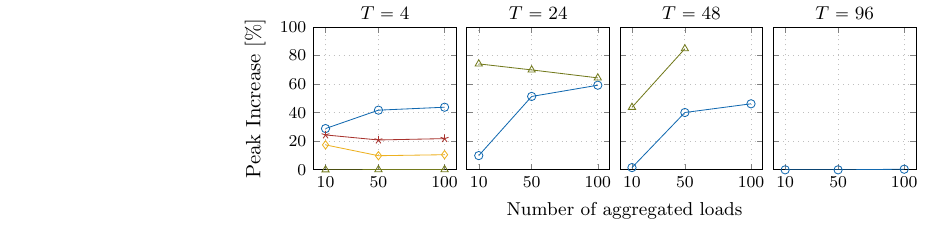}
    \caption{Evolution of the total computation time (top) and the aggregated peak increase (bottom) for different horizon lengths and aggregated load numbers.}
    \label{fig:all_methods_peak}
\end{figure}

The method developed in this paper is based on the physical intuition of a worst-case energy dispatch. As such, it distinguishes from existing methods that mainly rely on geometric approximations of the aggregated polytope. Figs.~\ref{fig:all_methods_cost} and~\ref{fig:all_methods_peak} compare our method to existing ones when employed for cost and peak-minimization, respectively. Both figures consistently depict the median total computation time among all scenarios introduced in Section~\ref{sec:caseStudy}, as well as the median increase in energy cost or power peak, for different numbers of aggregated loads and horizon lengths. For all methods and for a given load number and horizon length, each scenario must be computed in less than 15~minutes, otherwise the method is considered too computationally intensive and discarded. For better reading, Figs.~\ref{fig:all_methods_cost} and~\ref{fig:all_methods_peak} also only display increases up to 100\%. 

When minimizing cost, only the zonotope approximation outperforms our method in terms of accuracy, for a one-hour horizon. However, as the horizon length increases, our method results in a 5\%-cost increase, while other methods consistently display a higher cost increase and become less accurate. A higher cost increase is observed for a 12-hour horizon. By shifting the energy demand of loads over a half-day, we can profit from mid-day low prices, resulting in a low total cost. As such, the relative increase in cost appears as larger for this horizon.  

However, Fig.~\ref{fig:all_methods_peak} indicates a large peak increase when using our method compared to the unaggregated case. For a one-hour horizon, all existing approximations outperform our method. Notably, the zonotope approximation performs very well. Yet, when increasing the horizon length, all approximation methods become less accurate. Only our method performs equally when increasing the horizon length. Additionally, a surprisingly low peak increase is observed for a 24-hour horizon. Over a day, the inflexible household demand's peak occur in the evening. By shifting loads' energy demand throughout the day, the peak demand may remain close to the evening peak, leading to a small peak increase compared to the unaggregated case. 

The better performances of our method when minimizing cost instead of peak relates to its nature. It assumes a worst-case energy dispatch characterized by alternate maximum and minimum power consumptions. Hence, our method is particularly accurate in approximating the range of feasible aggregated energy values in such a case. When minimizing cost, loads alternate between minimum and maximum power consumptions, in order to profit from low prices, explaining the better cost performances. Additionally, we believe that better approximations of the lower conservative bound than the linear one can significantly improve the peak power performances of our method. 

Our method not only provides a stable accuracy across time horizons, but it is also computationally efficient. As Figs.~\ref{fig:all_methods_cost} and~\ref{fig:all_methods_peak} depict, while all existing methods are computationally intensive, our method offers a comparable total computation time to the unaggregated case, and is even faster when computing the lowest peak power. But, unlike the unaggregated case, it only requires aggregators to share aggregated constraints on their energy flexibility potential, masking individual loads' information. While no existing method can compute an approximation of the aggregated polytope of 100~loads over one day in less than 15~minutes, our method can do so within a few seconds.

\subsection{Scalability}
\begin{figure}
    \begin{subfigure}[b]{0.9\columnwidth}
        \centering
        \hspace{-2.3cm}\includegraphics[width=1.3\columnwidth]{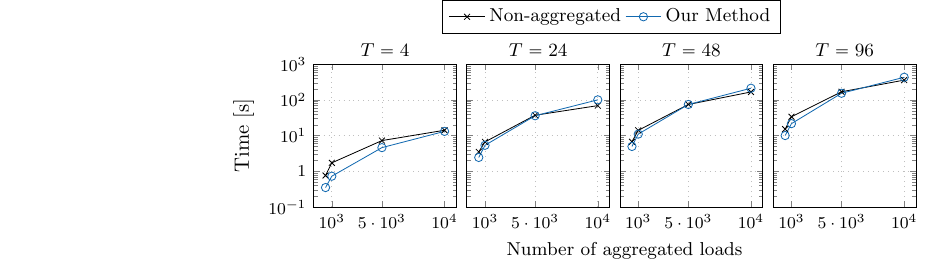}
        \vspace{-0.5cm}
        \caption{Cost Minimization}
    \end{subfigure}%
    \vspace{0.1cm}
    \begin{subfigure}[b]{0.9\columnwidth}
        \centering
        \hspace{-2.3cm}\includegraphics[width=1.3\columnwidth]{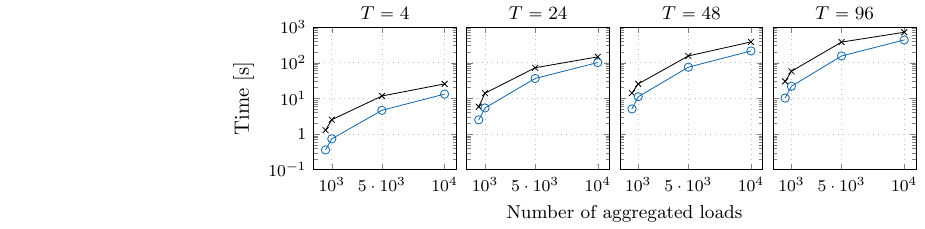}
        \vspace{-0.5cm}
        \caption{Peak Minimization}
    \end{subfigure}
    \caption{Evolution of the total computation time for different horizon lengths and large aggregated load numbers.}
    \label{fig:large_time}
\end{figure}

\begin{figure}
    \centering
    \hspace{-1.7cm}\includegraphics[width=0.85\columnwidth]{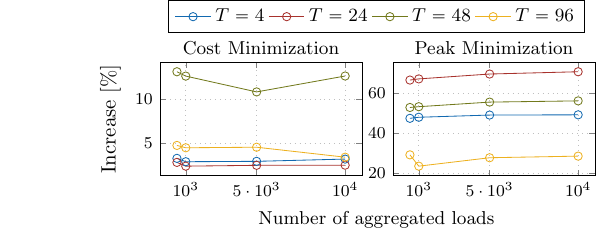}
    \caption{Evolution of the total computation time of the cost (left) and peak (right) minimization for different horizon lengths and large aggregated load numbers.}
    \vspace{-0.2cm}
    \label{fig:large_obj}
\end{figure}

As opposed to existing approximations, our method can compute an approximation of a polytope of many aggregated loads efficiently. Figs.~\ref{fig:large_time} and~\ref{fig:large_obj} assess our method's performances for realistic aggregated load numbers, from 500 up to 10,000~loads. As Fig.~\ref{fig:large_time} indicates, our method computes the aggregated lowest cost as fast as the unaggregated optimization and is even consistently faster in computing peaks than the unaggregated optimization while preserving individual loads' privacy. Furthermore, Fig.~\ref{fig:large_obj} depicts the increase in cost and peak, for large load numbers and horizon lengths. Despite the increasing load number, both increases remain steady for a given horizon length. 
The differences observed among horizon lengths can be explained by inflexible loads' peak and evening high prices, similarly to the differences observed in Figs.~\ref{fig:all_methods_cost} and~\ref{fig:all_methods_peak}.

\subsection{Feasibility}
\begin{figure}
    \centering
    \hspace{-1cm}\includegraphics[width=0.93\columnwidth]{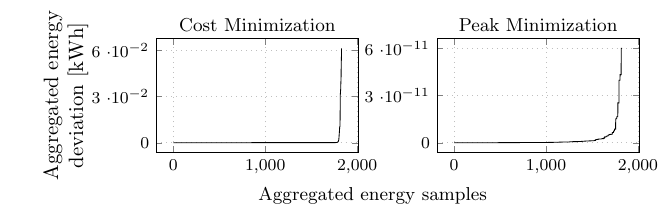}
    \caption{Cumulative root-mean squared between the requested and delivered aggregated energy.}
    \label{fig:feasibility}
\end{figure}

An inner approximation of the aggregated flexibility limits the range of aggregated energy, so that any aggregated energy value can be decomposed into feasible individual energy consumptions. If an aggregated request cannot be disaggregated, the delivered aggregated signal deviates from the request, so as to guarantee loads' power and energy constraints. Fig.~\ref{fig:feasibility} displays the cumulative root-mean squared error between the requested and delivered aggregated signal, for all scenarios generated in the scalability study, i.e., for load numbers between 500 and 10,000 and one-hour to one-day long horizons. We observe very small deviations for both cost and peak minimization, especially as compared to the number of aggregated loads, confirming that our method is an inner approximation.

\section{Conclusion}
\label{sec:Conclusion}
This paper introduces a novel approximation of the aggregated flexibility of power and energy-constrained flexible loads. Using the concept of a worst-case energy dispatch, we develop a conservative upper and lower bound that constrain the aggregated energy consumption only based on the previous aggregated energy consumption, as a piecewise linear function of this previous energy value. As our approximation may be non-convex, we additionally suggest a linear approximation of the piecewise linear bounds. 

Based on a comparative case study, we demonstrate that our method outperforms existing approximation techniques not only in terms of computational complexity but also in terms of accuracy. Besides, our method's accuracy remains steady, even for increasing load numbers.

Although the presented method shows promising results, future works should refine the convex approximation of the piecewise linear bounds, specifically for the lower one. Further studies should also extend the methodology to include more loads. Specifically, they should seek to extend it to time-varying power ratings that can be negative.

\bibliographystyle{ieeetr}
\bibliography{reference.bib}

\end{document}